\documentclass[12pt]{article}

\def\half{{1\over2}}

\def\bb#1{\hbox{\mybb#1}}

\def\be{ \begin{equation}}
\def\ee{ \end{equation}}
\def\N{\nabla}
\def\bar{\overline}
\def\tilde{\widetilde}
\def\e{\epsilon}
\def\Nb{\bar \N}
\def\t{\theta}
\def\bt{\bar\theta}
\def\s{\sigma}
\def\ba{\bar a}
\def\bb{\bar b}
\def\bc{\bar c}
\def\bd{\bar d}

\def\bg{\bar g}
\def\bG{\bar G}
\def\bh{\bar h}
\def\bj{\bar j}

\def\bR{\bar R}

\def\half{1\over 2}
\def\W{\mathcal W}
\def\Y{\mathcal Y}

\begin{document}
\thispagestyle{empty}
{\baselineskip=12pt
\hfill IFP-0002-UNC 

%\hfill hep-th/00

\hfill February 2000 

\vspace{1.0cm}}
\begin{center}
\large \textbf {Vertex operators in $AdS_3\times S^3$ Background with NS-NS 
Flux using Berkovits-Vafa-Witten Variables}
\end{center}

\bigskip
\bigskip
\centerline{\large  M. Langham\footnote{Email: mlangham@physics.unc.edu}}
\medskip
\centerline{\it Department of Physics and Astronomy}
\centerline{\it University of North Carolina, Chapel Hill, NC 27599-3255}
\bigskip
\bigskip
\bigskip

\vskip 1.0 truein
\parindent=1 cm

\begin{abstract}
\noindent
String equations for the vertex operators of $type~IIB$ on $AdS_3 \times 
S^3$ background with NS-NS flux are calculated using
Berkovits-Vafa-Witten formalism.  With suitable field definitions, the
linearized field equations for six-dimensional supergravity and a tensor 
multiplet on $AdS_3 \times S^3$ are recovered from these.  We also discuss 
the massless degrees of freedom that survive the $S^3$ Kaluza-Klein 
compactification and how our vertex operators are related to the vertex 
operators introduced by Giveon, Kutasov, and Seiberg.
\end{abstract}

\vfil\eject
\setcounter{page}{1}
\section{Introduction}
\quad The problem of string quantization with Ramond-Ramond flux has received more attention due to the AdS/CFT correspondence\cite{mal}.  The construction of the worldsheet superconformal field theory of $type~IIB$ on $AdS_3\times S^3$ with Ramond-Ramond flux\cite{BVW} makes use of the $N = 4$ topological 
string formulated by Berkovits and Vafa\cite{BV}. The Berkovits-Vafa quantization is a hybrid of Ramond-Neveu-Schwarz\cite{RNS} and Green-Schwarz\cite{GS} formalisms.  The spacetime part is constructed from GS-like variables and the compactification variables are in RNS formalism.  The supersymmetry is made manifest using GS-like variables and the supercharges do not have a singular OPE with the compactification variables.

The purpose of \cite{BVW} was to address string quantization
 with R-R flux, but $AdS_3\times S^3$ background with NS-NS flux is also 
considered in the paper.  In this paper we compute the constraint equations 
for the vertex operator for this simpler theory to make contact with earlier 
literature concerning NS-NS flux (e.g.\cite{balog, petro, moha, hwang, hwang1, bars, satoh,T,Gab,GKS,Leigh,oleg,peso}).  Also, see \cite{petro1} and \cite{egp1} for brief history of string theory in $AdS_3$. 

$AdS_3 \times S^3$ background 
with NS-NS flux is an exact background whose conformal field theory is a WZW 
model with target
space $SU'(2\mid 2)$. The WZW model on $SU'(2\mid 2)$ is a special limit of 
a two parameter conformal field theory\cite{BVW}.  The two parameters are the 
curvature of the target space and the level of the affine Kac-Moody algebra 
associated with the supergroup.  The generators of $SU'(2\mid 2)$ are then 
incorporated in a straightforward way in constructing the $N=4$ topological 
superconformal algebra.  The $N=4$ algebra is then used to impose physical 
and gauge conditions on vertex operators to  
obtain supersymmetric string constraints leading to string equations 
of motion. 
As in \cite{witten}, we will show the eqiuvalence of the string equations 
and the supergravity field equations in the $AdS_3 \times S^3$ background, 
but with NS-NS flux and different field identifications.  This differs from the RNS formalism, when one has to calculate various
amplitudes and then derive the form of the Lagrangian density to obtain
the supergravity field equations. 

In section two and three, we calculate supersymmetric string constraint 
equations for the spacetime vertex operators.  The constraint equations are 
used to solve for the vertex operator.  We then compute the string equations 
of motion and gauge conditions for the physical degrees of freedom carried 
by the 
vertex operator.  In section four, we linearize $N=(0,2), D=6$ supergravity 
with one tensor multiplet around the $AdS_3\times S^3$ metric and 
self-dual three form field strength.  
With suitable field definitions we show that the string and supergravity 
equations of motion match.  In the final section, 
we discuss the degrees of freedom remaining massless under the 
Kaluza-Klein compactification on $S^3$ and how they can be represented by 
vertex 
operators introduced by Giveon, Kutasov and Seiberg\cite{GKS}.

\section{$N=4$ Algebra and Supersymmetric String Constraint Equations}
\quad The topological $N=4$ 
superconformal algebra(SCA) is the key ingredient for computing the 
supersymmetric string constraints 
for the compactification independent massless vertex operator for the $type~
IIB$ on $AdS_3\times S^3$ background with NS-NS flux.  $N=4$ algebra can be 
constructed in the following 
way.  One takes the critical $N=1$ RNS superconformal field theory and 
constructs $N=2, c=6$ SCA by combining the matter and the ghosts.  For 
example, 
$$T = T_{N=1} + {1\over 2}\partial (bc + \xi\eta) = T_{N=1} - {1\over 2}
\partial J,$$
$$G^+ = J_{BRST} + {\partial}^2c + \partial (c\xi\eta),$$
$$G^- = b,$$
$$J = cb + \eta\xi,$$
where $T_{N=1}$ is the combined energy-momentum tensor of the original $N=1$
matter and ghost fields and the $(\beta,\gamma)$ are the 
super-reparameterization 
ghosts bosonized as ($\beta = e^{-\phi}\partial\xi, 
\gamma = e^{\phi}\eta).$  For $N=2$ SCA, $c=6$ is the critical dimension 
allowing the $N=2$ string to be reformulated as an $N=4$ string. For instance, if $J$ is the $U(1)$ 
current of $N=2, c=6$ SCA, then the three $SU(2)$ currents are $J, 
e^{-\int J}$
 and $e^{+\int J}$.  The four supercurrents are $G^{\pm}, {\tilde G}^- = 
[e^{-\int J}, G^+]$ and ${\tilde G}^+ = [e^{+\int J}, G^-].$  Normally, 
the four supercurrents $G^{\pm}$ and 
${\tilde G}^{\pm}$
 in $N=4$ SCA have conformal dimensions of ${3\over 2}$ and are charged under 
the $U(1)$ current.  In the topological
 string the the $U(1)$  current $J$ is constructed out of ghosts and the 
stress-energy tensor is modified 
or twisted by $-{\half}\partial J$.  Then the conformal dimensions of the 
generators are shifted by $-{\half}$ of their $U(1)$ charge.  The conformal 
dimensions of $G^+$ and ${\tilde G}^+$ are then one and their zero modes are 
used as 
``BRST'' charges.  The $N=4$ topological SCA for the $SU'(2\mid 2)$ target 
space is given by\cite {BVW},
$$T = -T_{SU'(2\mid 2)} - {1\over 2}\partial\rho\partial\rho - {1\over 2}
\partial\s\partial\s + {3\over 2}{\partial}^2(\rho + i\s) + T_C,$$
$$G^+ = -e^{-2\rho-i\s}(p)^4 + {i\over 2}e^{-\rho}(p_ap_bK^{ab} + 
{{p_a\partial p^a}\over {2k}})$$
$$-e^{i\s}[T_{SU'(2\mid 2)} + {1\over 2}\partial (\rho + i\s)\partial
 (\rho + i\s) -{1\over 2}{\partial}^2(\rho + i\s)] + G^+_C, $$
$$G^- = e^{-i\s} + G^-_C,$$
$${\tilde G}^+ = e^{iH_C + \rho} + e^{\rho + i\s}{\tilde G}^+_C,$$
$${\tilde G}^- = e^{-iH_C}\{-e^{-3\rho-i2\s}(p)^4 + {i\over 2}e^{-2\rho - i\s}
(p_ap_bK^{ab} + {{p_a\partial p^a}\over {2k}})$$
$$-e^{-\rho}[T_{SU'(2\mid 2)} + {1\over 2}\partial (\rho + i\s)\partial
 (\rho + i\s) -{1\over 2}{\partial}^2(\rho + i\s)]\} + e^{-\rho -i\s}
{\tilde G}^+_C,$$
$$J = \partial (\rho + i\s) + J_C,$$
$$J^{++} = e^{\rho + i\s}J^{++}_C,$$
$$J^{--} = e^{-\rho -i\s}J^{--}_C.$$
$T_{SU'(2\mid 2)}$ is the stress tensor for the $SU'(2\mid 2)$ WZW model, 
$$T_{SU'(2\mid 2)} = {1\over {8k}}\epsilon_{abcd}K^{ab}K^{cd} - 
{1\over {2k}}\epsilon_{\alpha\beta}S^{a\alpha}S_a^{\beta}$$
$$= {1\over {8k}}\epsilon_{abcd}j^{ab}j^{cd} + p^a\partial\theta_a.$$

 The cohomology of ${\tilde G}^+_0$ is taken
to be trivial so that the physical vertex operator $\Phi^+$ is written
in terms of $U(1)$ neutral vertex operator $V$, $\Phi^+ = {\tilde G}^+_0V$
($O_n\Phi$ denotes $(h+n)$th pole in the OPE between the operator $O$ of 
conformal dimension $h$ and $\Phi$).  
$\Phi^+$ is BRST closed under $G_0^+$ because the OPE $G^+{\tilde G}^+$ is 
nonsingular\cite{BV}.  
 The gauge invariance of $V$ follows from the fact that $V$ is defined up to 
 the equivalence class of the ``BRST'' charges.  The gauge is fixed by 
choosing
$$G^-_0V = {\tilde G}^-_0V = 0.$$
Furthermore, the gauge-fixing condition and $G^+_0\Phi^{+} = 0$ imply
$T_0V = 0.$  Similar conditions also arise in the right-moving sector.  
In summary, we have the following physical and gauge conditions which 
$V$ must satisfy
(the right movers are barred):
\begin{eqnarray} T_0V = {\bar T}_0V = 0, \nonumber \\
G^-_0V = {\tilde G}^-_0V = {\bar G}^-_0V = \bar {\tilde G}^-_0V = 0\nonumber
\\
G^+_0{\tilde G}^+_0V = {\bar G}^+_0\bar {\tilde G}^+_0V = 0.
\end{eqnarray}
$V$ is used to describe the most general massless vertex operators that are independent of compactification fields.  It is
 \be
 V=\sum_{m,n =-\infty}^{+\infty} e^{m(i\sigma+\rho)+n(i\bar\sigma+\bar\rho)}
V_{m,n} (x,\theta,\bar\theta). 
\ee
$G^-_0 V =\bar G^-_0 V=0$
implies that 
$V_{m,n}=0$ for $m>1$ or  $n>1$.
$\tilde G^-_0 V=\bar{\tilde G}^-_0 V =0$ tells us that $V_{m,n}=0$ for
 $m<-1$ or $n<-1.$  This leaves 
$$V_{1,1}, V_{1,0}, V_{1,-1},$$
$$V_{0,1}, V_{0,0}, V_{0,-1},$$
$$V_{-1,1}, V_{-1,0}, V_{-1,-1}$$ to consider.

$\tilde G^-_0 V=\bar{\tilde G}^-_0 V =0$ gives the following conditions on 
$V_{m,n}(m,n = 1,0,-1)$:

\be
\N^4 V_{1,n} = 0,
\ee
\be
{1\over 6}\e^{abcd}\N_b\N_c\N_d V_{1,n} = 
 i({j_0}^{ab} - {1\over 4k}{\delta}^{ab})\N_b V_{0,n},
\ee
\be
{1\over 2}\e^{abcd}\N_c\N_d V_{o,n} = ij_0^{ab} V_{-1,n},
\ee
\be
{j_0}^{ab}\N_a \N_b V_{1,n} = 0,
\ee
\be
\N_a V_{-1,n} = 0,
\ee
and
$$  $$
\be
{\bar \N^4} V_{m,1} = 0,
\ee
\be
{1\over 6}\e^{\bar a\bar b\bar c\bar d}\bar\N_{\bar b}\bar\N_{\bar c}\bar\N_{\bar d}
 V_{m,1} = i ({j_0}^{\bar a\bar b} - {1 \over 4k}\delta^{\bar a\bar b})
\Nb_{\bar b} V_{m,0},
\ee
\be 
{1\over 2}\e^{\bar a\bar b\bar c\bar d}\bar\N_{\bar c}\bar\N_{\bar d}
V_{m,0} =ij_0^{{\ba} {\bb}}V_{m,-1},
\ee
\be
{j_0}^{\bar a\bar b}\Nb_{\bar a}\Nb_{\bar b} V_{m,1} = 0,
\ee
\be
\Nb_{\bar a} V_{m,-1} = 0.
\ee

Ansatz for $V_{1,1}$ is provided by \cite{BVW},
\begin{eqnarray}
V_{1,1}=\t^a\bar\t^{\bar a} V^{--}_{a\bar a}+
\t^a\t^b\bar\t^{\bar a}\s^m_{ab} \bar\xi^-_{m~\bar a}+
\t^a\bar\t^{\bar a} \bar\t^{\bar b}\s^m_{\bar a\bar b} \xi^-_{m~ a}\nonumber \\
+
\t^a\t^b\bar\t^{\bar a} \bar\t^{\bar b}\s^m_{ab}\s^n_{\bar a\bar b}
( g_{mn}+b_{mn}+\phi {\bar g}_{mn}) +
\t^a(\bar\t^3)_{\bar a} A^{-+~\bar a}_{ a}
+(\t^3)_a\bar\t^{\bar a} A^{+-~ a}_{\bar a}\nonumber  \\
+\t^a\t^b(\bar\t^3)_{\bar a}\s^m_{ab} \bar\chi_m^{+~\bar a}+
(\t^3)^a\bar\t^{\bar a} \bar\t^{\bar b}\s^m_{\bar a\bar b} \chi_m^{+~ a}+
(\t^3)_a(\bar\t^3)_{\bar a} F^{++~ a\bar a} .
\end{eqnarray}

The remaining eight superfields can be computed up to integration 
constants using the constraints (3) - (13).  These will contain new component 
fields, but these can be gauged away by the following gauge transformation:
$$\delta V = G^+_0\Lambda +\bar{G}^+_0\bar\Lambda
 +\tilde G^+_0\tilde\Lambda +\bar{\tilde G}^+_0\bar{\tilde\Lambda}.$$
The gauge parameters $\Lambda, {\tilde\Lambda}, \bar{\Lambda}$, and 
$\bar{\tilde\Lambda}$ are not arbitrary.  
$\Lambda$ and ${\tilde\Lambda}$ must be  annhilated by 
$\bar G^+_0\bar{\tilde G}^+_0$, 
$\bar\Lambda$ and $\bar{\tilde\Lambda}$  annihilated by 
$G^+_0\tilde G^+_0$, and all the gauge parameters are annihilated 
 by $T_0$, $\bar T_0$,
$G^-_0$, $\tilde G^-_0$, $\bar G^-_0$, and  $\bar {\tilde G}^-_0$.
The parameters are chosen as follows, 
$$\Lambda= e^{2\rho+i\s +n(\bar\rho+i\bar\s)} \lambda_n(x,\t,\bar\t),\quad
\bar{\Lambda}= e^{2\bar\rho+i\bar\s +n(\rho + i\s)}
 \bar{\lambda}_n(x,\t,\bar\t),$$
$$\tilde
\Lambda= e^{-\rho-iH_C^{GS}+n(\bar\rho+i\bar\s)}\tilde 
\lambda_n(x,\t,\bar\t) + {\tilde G}_0^-\bar{\tilde G}_0^+\bar{\tilde G}_0^-
{\hat \Lambda},$$
$$\bar{\tilde \Lambda}= e^{-\bar\rho-i\bar H_C^{GS}+n(\rho+i\s)} \bar
{\tilde\lambda}_n(x,\t,\bar\t) + {\tilde G}_0^-\bar{\tilde G}_0^+\bar{\tilde G}_0^-
{\hat \Lambda},$$
and $\hat{\Lambda}$ is defined as
$-{1\over 2}e^{(\rho + i\sigma) + (\bar\rho + i\bar\sigma)}\hat{\lambda}.$  In short, $V_{1,1}$ carries all the degrees of freedom and we will show that 
$V_{1,1}$ describes $D=6, N=(0,2)$ supergravity with one tensor multiplet.

Next we apply (6) and (11) on (13) to obtain gauge conditions on the 
fields.  They are
\be j^{ab}_0(\sigma^m_{ab}{\bar\xi}^-_{m~\ba}) = 0, \ee
\be j^{ab}_0(\sigma^m_{ab}\sigma^n_{\ba\bb}G_{mn}) = 0,\ee
\be j^{ab}_0(\sigma^m_{ab}{\bar\chi}^{+-~\ba}_m) = 0,\ee
\be \epsilon_{abcd}j^{cd}_0A^{+-~a}_{\ba} = 0,\ee
\be \epsilon_{abcd}j^{cd}_0(\sigma^m_{\ba\bb}\chi^{+~a}_m) = 0,\ee
\be \epsilon_{abcd}j^{cd}_0F^{++ ~a\ba} = 0.\ee
$$ $$
\be j^{\ba\bb}_0(\sigma^m_{\ba\bb}\xi^-_{m~a}) = 0,\ee
\be j^{\ba\bb}_0(\sigma^m_{ab}\sigma^n_{\ba\bb}G_{mn}) = 0,\ee
\be j^{\ba\bb}_0(\sigma^m_{\ba\bb}\chi^{+~a}_m) = 0, \ee
\be \epsilon_{\ba\bb\bc\bd}j^{\bc\bd}_0A^{-+~\ba}_a = 0,\ee
\be \epsilon_{\ba\bb\bc\bd}j^{\bc\bd}_0(\sigma^m_{ab}{\bar\chi}^{+~\ba}_m)
 = 0,\ee
\be \epsilon_{\ba\bb\bc\bd}j^{\bc\bd}_0F^{++~a\ba} = 0,\ee
where $G_{mn}=g_{mn} + b_{mn} + {\bar g}_{mn}\phi$, and $\bg_{mn}$ is the metric of $AdS_3\times S^3$.  Now we use the definitions of invariant derivatives\cite{witten} on the $SO(4)$ group
 manifold to obtain the gauge conditions for the bosonic fluctuations.  
\be D^m(g_{mn} + {\bar g}_{mn}\phi) -{\half} Z_{npq}b^{pq} = 0,\ee
\be D^mb_{mn} = 0,\ee
\be \sigma^p_{ab}D_pA^{+-~a}_d -{\half} \epsilon_{abcd}A^{+-~ca} = 0,\ee
\be \sigma^p_{ab}D_pA^{-+~b}_d + {\half} \epsilon_{abcd}A^{-+~cb} = 0, \ee
\be 
\sigma^p_{ab}D_pF^{++~ae} -
 {\half} \epsilon_{abcd}\delta^{ed}F^{++~ac} = 0,
 \ee
\be
 \sigma^p_{ab}D_pF^{++~ea} -
 {\half} \epsilon_{abcd}\delta^{ed}F^{++~ac} = 0.
 \ee
Here $Z_{mnp} = (\s_m\s_n\s_p)_{ab}\delta^{ab}$ is the selfdual combination 
of sigma matrices.

\section{String Equations of Motion}

\quad The physical condition that gives rise to string equations is $T_0V = 
{\bar T}_0V = 0,$ or ${1 \over 8}\epsilon_{abcd}j^{ab}j^{cd}V_{1,1} = 
{1\over 8}\epsilon_{\ba\bb\bc\bd}j^{\ba\bb}j^{\bc\bd}V_{1,1} = 0.$
The string equations for the bosonic fields are
\begin{eqnarray}
D^pD_p(g_{mn} + b_{mn} + {\bar g}_{mn}\phi) =\nonumber \\
Y^q_{~(m}g_{n)q} - Y_{pmnq}g^{pq}
+ 2Z_{pq(n}D^pb_{~m)}^q + 2Z_{mnp}D^p\phi \nonumber \\
 + 2Y_{mn}\phi
-{3\over 2}Y^p_{~[m}b_{n]q} + 2 Z_{pq[n}D^pg_{~m]}^q,
\end{eqnarray}
\be
D^pD_pF^{++~ab} = 0
\ee
\be
D^pD_pA^{+-~a}_b - \s^p_{cb}D_pA^{+-~ca} = 0,\ee
\be
D^pD_pA^{+-~a}_b + \s^p_{cb}D_pA^{-+~ca} = 0,\ee
\be
D^pD_pV^{--}_{ab}+\delta^{cd}\s^p_{bc}D_pV^{--}_{ad}
+\delta^{cd}\s^p_{ca}D_pV^{--}_{db} -{1\over 2}\epsilon_{ab}^{~~cd}V^{--}_{cd}
=0,\ee
where $Y^{mn} \equiv \s^m_{ab}\s^n_{cd}\delta^{ac}\delta^{bd}$, and 
$$Y_{mnpq} \equiv {1\over 4}(\bg_{mp}Y_{nq} + \bg_{nq}Y_{mp} \\
-\bg_{np}Y_{mq} -\bg_{mq}Y_{np}).$$  
$\bg_{mn} = {1\over 2}\s_m^{ab}\s_{n~ab}$ follows from the algebra of sigma 
matrices, $\s_m^{ab}\s_{n~ad} + \s_n^{ab}\s_{m~ad} = 
\bg_{mn}\delta^b_{~d}.$  Furthermore, we have used the following 
identity, \begin{eqnarray} 
Y^{mnpq} = {1\over 4}\delta^{ah}\delta^{kg}[\s^m_{ka}\s^n_{ge}\s^p_{hf}
+ \s^m_{ak}\s^n_{he}\s^p_{gf} + \s^m_{kf}\s^n_{ga}\s^p_{he} + 
\s^m_{af}\s^n_{hk}\s^p_{ge}]\s^{q~ef}.
\end{eqnarray}

\section{Linearized Supergravity Field Equations in $AdS_3\times ~S^3$}

\quad In the bosonic sector, the $N=(0,2), D=6$ supergravity field equations 
are
\cite{Romans, Sezgin} 
\be R_{mn} = H^i_{mpq}H^{i~pq}_n + K_{mpq}K^{~~pq}_n + D_m\phi^iD_n\phi^i, \ee
\be D^mD_m\phi^i = {2 \over 3}H^{i~mnp}K_{mnp}, \ee
\be H^i_{mnp} = {1 \over {3!}}e_{mnpqrs}H^{i~qrs}, \qquad
K_{mnp} = -{1 \over {3!}}e_{mnpqrs}K^{qrs},\ee
where $e_{mnpqrs} \equiv {1 \over {\sqrt{-g}}}g_{mm'}...g_{ss'}
\epsilon^{m'n'p'q'r's'}$, and $\epsilon^{012345} = 1, 1\leq i \leq 5.$
The Ricci tensor is defined as $R_{mn} = g^{pq}R_{pmqn}$.  In $AdS_3\times 
S^3$ background the following relation holds, $\bR_{pmqn} = {1\over 4}
(\bg_{pq}\bR_{mn} + \bg_{mn}\bR_{pq} - \bg_{mq}\bR_{pn} - \bg_{pn}\bR_{mq}).$
Fields in (38)-(40) will be linearized around a background with 
non-vanishing $AdS_3 \times S^3$ metric $\bg_{mn}$ and selfdual three form 
field 
strength $\bG^5_{mnp}$, i.e. only one of the components of the selfdual 
tensor 
is non-zero.  The fluctuations about this background are expressed
as\cite{Sezgin} \begin{eqnarray}
g_{mn} = \bg_{mn} + h_{mn}, \quad H^i_{mnp} = \bG^i_{mnp} + g^i_{mnp},
 \\ K_{mnp} ={\bar K}_{mnp} + g^6_{mnp} + \bG^j_{mnp}\phi^j. \end{eqnarray}
From the scalar equation, the zeroth order is given by
 $\bG^i_{mnp}{\bar K}^{mnp} = 0$.  To satisfy this we choose ${\bar K}_{mnp}
= 0$.  From the Einstein equation, we
get the identity $\bR_{mn} = \bG_{mpq}\bG_n^{~pq}$, where
$\bG_{mnp}\equiv \bG^5_{mnp}$.

The linearized bosonic field equations that obey the zeroth order
are
\be
{1 \over {3!}}{\bar e}_{mnp}^{~~~~qrs}g^i_{qrs} = g^i_{mnp} -
 3h_{q[m}\bG^{i~~~q}_{np]}
 + {1 \over 2}h^q_{~q}\bG^i_{mnp},\ee
\be {1 \over {3!}}{\bar e}_{mnp}^{~~~~~qrs}g^6_{qrs} = 
-g^6_{mnp} - 2\phi^i\bG^i_{mnp},  \ee
\be D^pD_p\phi^i = {2\over 3}\bG^{i~mnp}g^6_{mnp}, \ee 
 and 
\begin{eqnarray}
D^pD_ph_{mn} = 2\bR^q_{~(m}h_{n)q} + 2\bR_{pmnq}h^{pq} 
+2D_{(m}D^ph_{n)p}\nonumber \\ - D_mD_nh^q_{~q} - 4\bG_{pq(m}g^{5~~pq}_{n)}
+4\bG_{mp}^{~~~q}\bG_n^{~~pr}h_{qr}. \end{eqnarray}
$g^i_{mnp} = 3D_{[m}b^i_{np]}$ and $g^6_{mnp}=3D_{[m}b^6_{np]}$ are no 
longer selfdual and anti-selfdual, 
respectively, 
but they are exact.

To obtain second order field equations for the tensors we take a 
covariant divergence of (43) and (44). We write the NS-NS two form as $b_{mn}
 = Ab^6_{mn} + Bb^5_{mn}$ rather than the most general form $Ab^6_{mn} + 
B^ib^i_{mn}$ because we will have to set $B^x = 0$ for $x= 1,2,3,4$ to 
satisfy (32).
 Since the covariant divergence
of the dual of an exact form vanishes, we get
\begin{eqnarray}
D^pD_pb_{mn} =  \bR^s_{~~[m}b_{n]s} - 2AD^p\phi^5\bG_{pmn} \nonumber \\
+ {B \over C}\bG_{mnp}D_qg^{pq} + 
{2B \over C}D^pg_{q[m}\bG_{n]p}^{~~~q} - 2D_{[m}D^pb_{n]p}, 
\end{eqnarray}
and \be D^pD_pb^x_{mn} = -\bR^p_{~~[m}b^x_{n]p}. \ee

   We have expanded the fluctation $h_{mn}$ in terms of the graviton $g_{mn}$,
 $h_{mn} = {1 \over C}g_{mn} + {1 \over 6}\bg_{mn}h^q_{~q}$, for some constant
C to be determined.  Next we use the gauge conditions (26) and (27).  (47) 
then becomes 
\begin{eqnarray}
D^pD_pb_{mn} = -\bR^p_{~~[m}b_{n]p} - 2AD^p\phi^5\bG_{mnp} \nonumber \\
+ {B\over C}(-D_q\phi + {1\over 2}Z_{qpr})\bG_{mn}^{~~~~q} + 
{{2B}\over C}D^pg_{q[m}\bG_{n]p}^{~~~~q}. \end{eqnarray}  
Comparing (49) with the antisymmetric part of (32) yields the following 
identifications: 
$\bG_{mnp} = {C\over B}Z_{mnp}$, $\phi = {{2AC}\over {3B}}\phi^5$  
and $\bR_{mn} = {1\over 2}Y_{mn}$, using $Z^{mn}_{~~~r}Z^{rpq}b_{pq} = 
-2Y^{q[m}b^{n]}_{~~~q}.$

The scalar equation (39) reduces \be
D^pD_p\phi^5 = {2\over {3A}}\bG^{pqr}H_{pqr} - {B\over {AC}}\bR^{pq}g_{pq},\ee
and \be D^pD_p\phi^x = 0.\ee
We expand (46) by substituting $h_{mn} = {1\over C}g_{mn} + {1\over 6}
\bg_{mn}h^q_{~q}$.  We get
\begin{eqnarray}
D^pD_pg_{mn} + {C \over 6}\bg_{mn}D^pD_ph^q_{~q} = 
2\bR^q_{~~(m}g_{n)q} + 2\bR_{pmnq}g^{pq} + 2D_{(m}D^pg_{n)p}\nonumber \\
- {{2C}\over 3}D_mD_nh^q_{~q} - {{4C}\over B}\bG^{pq}_{~~~(m}H_{n)pq} 
+4\bG_{mp}^{~~~q}\bG_n^{~~pr}g_{qr} \nonumber \\
+ {{2C}\over 3}\bR_{mn}h^q_{~q} + {{4CA}\over B}
({1\over {6A}}\bg_{mn}\bG^{pqr}H_{pqr}
- {B\over {4AC}}\bg_{mn}\bR^{pq}g_{pq} - \bR_{mn}\phi^5), \end{eqnarray}
where $H_{mnp} = Ag^6_{mnp} + Bg^5_{mnp}.$ 
To reduce (52) further, we trace it to obtain an expression for 
$D^pD_ph^q_{~q}$, use the gauge condition (26) and use the identity 
$\bG_{mp}^{~~q}\bG_n^{~pr}g_{qr} = -\bR_{pmnq}g^{rq}$.

\quad Equation (52) will reduce to 
\begin{eqnarray}
D^pD_pg_{mn} - {1\over 5}\bg_{mn}D^pD_p\phi = -{1\over 30}Z_{pqr}H^{pqr}
\bg_{mn} - {4\over 5}\bg_{mn}\bR^{pq}g_{pq}\nonumber \\
+ 2\bR^q_{~~(m}g_{n)q} - 2\bR_{pmnq}g^{pq} - 
(2D_{(m}D_{n)}\phi + {{2C}\over 3}D_{(m}D_{n)}h^q_{~q})\nonumber \\
+ D_{(m}b^{pq}Z_{n)pq} - {{4C}\over B}D_{(m}b^{pq}\bG_{n)pq}
+ {{8C}\over B}\bG_{pq(m}D^pb_{n)}^{~~q} + {{2C}\over 3}\bR_{mn}
h^q_{~q}\nonumber \\
+ {{2C}\over {3B}}\bg_{mn}\bG^{pqr}H_{pqr} - {{4CA}\over B}\bR_{mn}\phi^5.
\end{eqnarray}
Next trace (53) to solve for $D^pD_p\phi$ and use it to put (52) in the same 
form as the symmetric part of (32).  The string equation 
does not have terms of the form $D_{(m}D_{n)}\phi$.  The simplest choice is 
to set $\phi = 
-{C\over 3}h^q_{~q}.$  To satisfy (32) we must also have $\bR_{mn} = 
{1\over 2}Y_{mn}$ and ${{2C}\over 3}h^q_{~q} - {{4CA}\over B}\phi^5 = 4\phi.$
  With $\phi = -{C\over 3}h^q_{~q}$, the latter expression gives 
$\phi = {{2CA}\over {3B}}\phi^5.$  This is the same expression we obtained 
from the tensor equation of motion.  Thus (53) will reduce to 
\begin{eqnarray}
D^pD_pg_{mn} + D^pD_p\phi = Y^q_{(m}g_{n)q} - g^{pq}Y_{pmnq}\nonumber \\
+ 2Y_{mn}\phi + Z_{pq(m}D_{n)}b^{pq}(1-{{4C^2}\over B^2})\nonumber \\
- {{8C^2}\over B^2}D_pb_{q(m}Z^{pq}_{~~~n)} + \bg_{mn}Z^{pqr}H_{pqr}
({{2C^2}\over {3B^2}} - {1\over 6}). \end{eqnarray}
We conclude that ${{C^2}\over B^2} = {1\over 4}$ or ${C\over B} = 
\pm{1\over 2}.$  For consistency we take ${C\over B} = {1\over 2}.$

In summary, the following identifications  
\begin{eqnarray}
\bg_{mn} = {1\over 2}\s_m^{ab}\s_{n~ab},\nonumber \\
\bR_{mn} = {\half}\s_m^{ab}\s_n^{cd}\delta_{ac}\delta_{bd},\nonumber \\
\bG^{mnp} = {\half}\s^m_{ab}\s^n_{cd}\s^{p~bd}\delta^{ac},\nonumber \\
B=2C, \qquad \phi = -{1\over 3}\phi^{\prime} = -{C\over 3}h^q_{~q}.
\end{eqnarray}\\
will allow the supergravity equation to be recovered from 
the string equation (32).
$$ $$

The supergravity equations of tensor and scalar fields, (48) and (51),
 are identified with the string equations (33)-(36) 
in the 
following way.  We know that the bispinor
$F^{++~ab}$ contains $H_{mnp}^{++}$ and a scalar $\phi^{++}$.  
In fact, we can use (30) and (31) to deduce
\be 
F^{++~ab} = \s^{p~ab}D_p\phi^{++} + (\s^m\s^n\s^p)^{(ab)}H^{++}_{mnp}
+ \delta^{ab}\phi^{++}, 
\ee
and 
\be D^pD_p\phi^{++} = 0. \ee
(52) with (30) or (31) implies $$D^pH^{++}_{pmn} = 0, \quad \mathrm{or}$$
\be D^pD_pb^{++}_{mn} = -\bR^p_{~~[m}b_{n]p}^{++}, \ee
  We can do the same for $b^{+-}_{mn}$ and $b^{-+}_{mn}$. 
Define \cite{witten} $F^{+-~a\ba} = 
(j^{\ba\bb} + \delta^{\ba\bb})A^{+-~a}_{\bb}$ and 
$F^{-+~a\ba} = (j^{ab} + \delta^{ab})A^{-+~\ba}_b$, then the gauge conditions
for $A^{+-~a}_{\ba}$ and $A^{-+~\ba}_a$ are satisfied if we write 
$$\epsilon_{abcd}j^{cd}F^{+-~a\ba} = 0,\quad 
\epsilon_{\ba\bb\bc\bd}j^{\bc\bd}F^{+-~a\ba} = 0, $$
$$\epsilon_{abcd}j^{cd}F^{-+~a\ba} = 0,\quad
\epsilon_{\ba\bb\bc\bd}j^{\bc\bd}F^{-+~a\ba} = 0.$$ 

These definitions provide us with equations for $b_{mn}^{+-}$, $\phi^{+-}$ and
$b_{mn}^{-+}$, $\phi^{-+}$ as in (56) and (57).  Now we are left with (36) 
to consider.  We define $F^{--~a\ba} = (j^{ab} + \delta^{ab})
(j^{\ba\bb} + \delta^{\ba\bb})V^{--}_{b\bb}$, then we can satisfy
$\epsilon_{abcd}j^{cd}F^{--~a\ba} = \epsilon_{\ba\bb\bc\bd}j^{\bc\bd}
F^{--~a\ba} = 0$ so that $\epsilon_{abcd}j^{ab}j^{cd}F^{--~a\ba} = 0.$  
Therefore, we can express the string equations (33)-(36) as
\begin{eqnarray} D^pD_pb^x_{mn} = -\bR^p_{~~[m}b^x_{n]p},\nonumber \\
D^pD_p\phi^x = 0, \end{eqnarray}
where $x = ++, +-, -+, \quad \mathrm{and} --$. (59) is identical
 to (48) and 
(51).

\section{The Massless Multiplets and the GKS Construction}

\quad  Following \cite{GKS}, numerous interesting papers 
on $type~IIB$ on
$AdS_3$ appeared
\cite{Leigh,OOguri,Seiberg, Kaz, Boer, Larsen}.  We summarize the basic things we 
need
to discuss the massless multiplet on $AdS_3$.
$$ $$   

$V_{1,1}$ contains degrees of freedom to describe $D=6, N=(0,2)$ supergravity 
and one tensor multiplet, with representations 
$$[(3,3) + 5(3,1) + 4(3,2)] + [(1,3) + 5(1,1) + 4(1,2)]$$
in $Spin(4) \sim SU(2)\times SU(2)$, the lightcone little group.

We denote the anti-selfdual tensor as $(1,3)$.  The Kaluza-Klein 
compactification to $AdS_3 \times S^3$ will leave us with 
\be (2,2) + 4(1,1) + 2(1,2) + 2(2,1)
\ee as massless propagating degrees of freedom, where the representation 
in (60) are under 
the $SO(4) \sim SU(2)\times SU(2)$ gauge group that comes
 from $S^3$.  The four massless degrees of freedom, $(2,2)$, comes from 
the linear combination of the anti-selfdual part of the NS-NS tensor and
 the dilaton.  The four moduli $4(1,1)$ are 
inherited from the four R-R scalars $\phi^x$.  The superpartners of the
 8 scalars transform as $2(1,2) + 2(2,1)$ under the gauge group.  These 
states come from the Kaluza-Klein reduction of the six dimensional 
fermions, $4(1,2).$  So intuitively, we can see how the Kaluza-Klein 
compactification of $type IIB$ on $AdS_3\times S^3\times M^4 (M^4 = T^4$
or $K3)$ gives $n$ massless mulitplet of (60);$n=5$ for $T^4$ and 
$n=21$ for $K3$.  As an example, $(2,2)$ come from 
$b^6_{mn}$ and $\phi$ of the $\t^a\t^b\bt^{\ba}\bt^{\bb}$ component of 
$V_{1,1}$ and $(2,2)$ is represented by $\phi_+^{(l+1,0)(l+1,0)}$ with $l=0$, 
(we use the results of \cite{Sezgin})
\begin{eqnarray}
\t^a\t^b\bt^{\ba}\bt^{\bb}[\s^{\mu}_{ab}\s^{\nu}_{\ba\bb}
({\bar e}_{\mu\nu\rho}
\sum_{l=0}^{\infty}{{l+2}\over {2(l+1)}}\partial^{\rho}\phi_+^{(l+1,0)
(l+1,0)}(x)Y^{(l+1,0)}(y) \nonumber \\
 + \bg_{\mu\nu}\sum_{l=0}^{\infty}{l\over {2(l+2)}}\phi_+^{(l+1,0)
(l+1,0)}(x)Y^{(l+1),0}(y)) + ...]
\end{eqnarray}
The greek indices label the coordinates of $AdS_3$.  The spherical harmonics 
of $S^3$ are represented by $Y^{(l_1,l_2)}$ and $(l_1,l_2)$ label the highest
weight of the gauge group $SO(4)$.  They are related to $SU(2)\times SU(2)$ 
isospin $(j,\bj)$ as $j={1\over 2}(l_1 + l_2)$ and $\bj = {1\over 2}
(l_1 - l_2)$, and $l_1 \geq l_2$ since $j,\bj = 0, {1\over 2}, 1, ...$
$$ $$

$\phi_+(x)$ is labeled by the $AdS_3$ energy $E_o$, helicity $s_o$ and 
$(l_1,l_2)$, i.e. $\phi_+^{(E_o,s_o)(l_1,l_2)}$.  
The $AdS_3$ energy $E_o$ and 
helicity $s_o$ are related to the spacetime conformal dimensions $h_{st} = 
{1\over 2}(E_o + s_o), \bh_{st} = {1\over 2}(E_o - s_o).$  $\phi_+$ is a 
chiral primary in the sense that the $SU(2)$ isospin is equal to the spacetime
 conformal dimension, $(j,\bj)= (h_{st},\bh_{st})$.

The massless multiplet (60) 
can be described in terms of vertex operators used by \cite{GKS,Leigh}.  
To be specific we take $Type~IIB$ on 
$AdS_3\times S^3\times T ^4$.  The holomorphic NS and its superpartner R 
vertex operators are
(see \cite{Leigh} for conventions)
\be {\W}_j = e^{-\phi}(\psi V_{jm})_{j-1}V'_{jm'}\ee
and
\be {\Y}_j = e^{-{\phi \over 2}}S^{\dot{A}}(SV_{jm}V'_{jm'})_{{j-{\half}},
j-{\half}}.\ee
To describe (60) we take $j={\half}$(see below).  In (62) and (63) we have used the following definitions:
 $$(\psi V_{jm})_{j-1} = (\psi^3V_{jm} - {\half}\psi^+V_{j~m-1}
 - {\half}\psi^-V_{j~m+1})_{j-1}$$
and
$$(SV_{jm}V'_{jm'})_{{j-{\half}},j-{\half}} =$$
$$S^{\dot{1}}V_{j~m-{\half}}V'_{j~m-{\half}} 
- S^{\dot{2}}V_{j~m-{\half}}V'_{j~m'+{\half}}$$
$$+ S^{\dot{3}}V_{j~m+{\half}}V'_{j~m'-{\half}} 
- S^{\dot{4}}V_{j~m+{\half}}V'_{j~m'+{\half}}.$$

$S^{\dot{A}}$, ${\dot {A}} = 1,2$, is the  spin field of $T^4$ and 
$S^{\dot{\alpha}}$, $\alpha = 1, ..., 4$, is the spin field of $AdS_3 \times
 S^3$ given (up to a cocycle) by $S^{\dot{\alpha}} = 
e^{{i\over 2}(\epsilon_1H_1 + \epsilon_2H_2 + \epsilon_3H_3)}.$ $\epsilon_i 
= \pm$ and $\epsilon_1\epsilon_2\epsilon_3 = -1.$  $S^\alpha$ is given by 
$\epsilon_1\epsilon_2\epsilon_3 = +1.$  We have chosen
$$S^{\dot{1}} = (\e_1, \e_2, \e_3) = (-,-,-),$$
$$S^{\dot{2}} = (-,+,+),$$
$$S^{\dot{3}} = (+,-,+),$$
$$S^{\dot{4}} = (+,+,-).$$

Setting $j=j'$ gives worldsheet conformal dimension of 1 for (62) and
(63). 
The subscript
 below the closed parenthesis of $(\psi V_{jm})_{j-1}$ means that it has $j-1$ 
representation under the $SU(1,1)$ Kac-Moody currents.  Similarly, 
$(SV_{jm}V'_{jm'})_{{j-{\half}},j-{\half}}$ has
 $(j-{\half}, j-{\half})$ representation under 
$SU(1,1)\times SU(2)$. (62) and (63) are BRST invariant and describe
 chiral primaries in spacetime because the spacetime conformal dimension
is equal to the $SU(2)$ quantum number; 
 $h_{st} = 
j_{SU(2)} = j_{SU(1,1)} + 1$ for the NS vertex operator and $h_{st} = 
j_{SU(2)} = j_{SU(1,1)} + {\half}$ for the R vertex operator.  $j_{SU(2)}$
 and $j_{SU(1,1)}$ refers to the total representation of the vertex operator
 under the Kac-Moody currents, i.e. $j_{SU(1,1)} = j-1, j_{SU(2)} = j$ for
(62), and $j_{SU(1,1)} = j_{SU(2)} = j-{1\over 2}$ for (63).  The 
$AdS_3$ part of NS vertex operators belong to the discrete unitary series 
$D^+_{\tilde j}$ with 
${\tilde j} = j_{SU(1,1)} = -{\half}, 0, {\half}, 1, {3\over 2}, ...$. The NS highest 
weight state is $j_{SU(1,1)} = -{\half}$ or $j = {\half}$ 
and the R highest weight state has $j_{SU(1,1)} = 0$.  Then the four NS-NS scalars are described by 
$$(j_{SU(2)}, {\bar j}_{SU(2)}) = ({\half},{\half}) = (h_{st},{\bar h}_{st}),$$
and the four R-R scalars are labelled with 
$$(j_{SU(2)}, {\bar j}_{SU(2)}) = (0,0) = (h_{st},{\bar h}_{st}).$$

\vfil\eject
   {\textbf {Acknowledgement:}} The author would like to thank Louise Dolan for many helpful discussions.

\end{document}